\documentclass[traditabstract]{aa}
\usepackage{txfonts}
\usepackage{graphicx}
\usepackage{natbib}
\usepackage{longtable,lscape}
\bibpunct{(}{)}{;}{a}{}{,}

\begin{document}

\title{The doubly eclipsing quintuple low-mass star system\\
  1SWASP J093010.78+533859.5}

\author{M.~E.~Lohr\inst{\ref{inst1}}\and
  A.~J.~Norton\inst{\ref{inst1}}\and E.~Gillen\inst{\ref{inst2}}\and
  R.~Busuttil\inst{\ref{inst1}}\and U.~C.~Kolb\inst{\ref{inst1}}\and
  S.~Aigrain\inst{\ref{inst2}}\and
  A.~McQuillan\inst{\ref{inst2}\and\ref{inst3}}\and
  S.~T.~Hodgkin\inst{\ref{inst4}}\and E.~Gonz\'{a}lez\inst{\ref{inst5}}}

\institute{Department of Physical Sciences, The Open University,
  Walton Hall, Milton Keynes MK7\,6AA, UK\\
  \email{Marcus.Lohr@open.ac.uk}\label{inst1}\and Sub-department of
  Astrophysics, Department of Physics, University of Oxford, Keble
  Road, Oxford OX1\,3RH, UK\label{inst2}\and School of Physics and
  Astronomy, Raymond and Beverly Sackler, Faculty of Exact Sciences,
  Tel Aviv University, 69978 Tel Aviv, Israel\label{inst3}\and
  Institute of Astronomy, Madingley Road, Cambridge CB3\,0HA,
  UK\label{inst4}\and Observatori Astron\'{o}mic de Mallorca, Cam\'{i}
  de l'Observatori s/n, 07144 Costitx, Mallorca\label{inst5}}
\date{Received 26 February 2015 / Accepted 17 April 2015}

\abstract {Our discovery of 1SWASP J093010.78+533859.5 as a probable
  doubly eclipsing quadruple system, containing a contact binary with
  $P\sim0.23$~d and a detached binary with $P\sim1.31$~d, was
  announced in 2013.  Subsequently, Koo et al. confirmed the detached
  binary spectroscopically, and identified a fifth set of static
  spectral lines at its location, corresponding to an additional
  non-eclipsing component of the system.  Here we present new
  spectroscopic and photometric observations, allowing confirmation of
  the contact binary and improved modelling of all four eclipsing
  components.  The detached binary is found to contain components of
  masses $0.837\pm0.008$ and $0.674\pm0.007$~M$_{\sun}$, with radii of
  $0.832\pm0.018$ and $0.669\pm0.018$~R$_{\sun}$ and effective
  temperatures of $5185_{-20}^{+25}$ and $4325_{-15}^{+20}$~K,
  respectively; the contact system has masses $0.86\pm0.02$ and
  $0.341\pm0.011$~M$_{\sun}$, radii of $0.79\pm0.04$ and
  $0.52\pm0.05$~R$_{\sun}$, respectively, and a common effective
  temperature of $4700\pm50$~K.  The fifth star is of similar
  temperature and spectral type to the primaries in the two binaries.
  Long-term photometric observations indicate the presence of a spot
  on one component of the detached binary, moving at an apparent rate
  of approximately one rotation every two years.  Both binaries have
  consistent system velocities around $-11$ to
  $-12$~km~s\textsuperscript{-1}, which match the average radial
  velocity of the fifth star; consistent distance estimates for both
  subsystems of $d=78\pm3$ and $d=73\pm4$~pc are also found, and, with
  some further assumptions, of $d=83\pm9$~pc for the fifth star.
  These findings strongly support the claim that both binaries -- and
  very probably all five stars -- are gravitationally bound in a
  single system.  The consistent angles of inclination found for the
  two binaries ($88.2\pm0.3$\degr and $86\pm4$\degr) may also indicate
  that they originally formed by fragmentation (around 9--10 Gyr ago)
  from a single protostellar disk, and subsequently remained in the
  same orbital plane.}

\titlerunning{The doubly eclipsing quintuple low-mass star system
  J093010}
\authorrunning{M.~E.~Lohr et al.}

\maketitle

\section{Introduction}

During the course of a search for orbital period variations in
short-period eclipsing binary candidates in the SuperWASP archive
\citep{pollacco}, described in \citet{lohr} and \citet{lohr13},
several unusual systems of particular astronomical interest were
encountered.  Amongst these, 1SWASP J234401.81\--212229.1, discussed in
\citet{lohr13b} and \citet{koen}, appears to be a triple system
containing an M+M dwarf contact binary.  Here we explore 1SWASP
J093010.78\-+533859.5 (hereafter J093010), an apparent quintuple
system containing two eclipsing binaries, one of only a handful of
known doubly eclipsing multiples.

Higher-order multiple systems are of value in general for testing
models of stellar formation and long-term dynamical stability.  When
they contain two double-lined spectroscopic and eclipsing binaries,
they provide a highly unusual opportunity to determine the component
stars' physical parameters and so to understand the structure and
potential origins of the whole system.

\section{Background}

A near neighbour of J093010 (1SWASP J093012.84\-+533859.6, hereafter
J093012) was identified as a candidate short-period eclipsing binary
in \citet{norton}, and its orbital period confirmed in \citet{lohr}.
However, in our subsequent more thorough search for eclipsing
candidates, J093010 was identified as exhibiting the same period and
light curve shape, but a higher mean flux and greater amplitude of
flux variability; hence, it was regarded in \citet{lohr13} as the
probable true source of the observed eclipsing variation, with J093012
being a fainter near-duplicate light curve, falling within the same
photometric aperture used by SuperWASP.  At the location of J093012
there is a magnitude 18 source listed in the USNO-B1 catalogue, separated
from J093010 by approximately 18\arcsec, which would not have been
detectable by SuperWASP in its own right.

The eclipsing source's apparently significant period change,
associated with an erratic O$-$C diagram, had been rejected initially
as the result of contamination by a nearby star \citep{lohr}, but in
\citet{lohr13} a fuller explanation was pursued.  Prior to analysis,
the data for J093010 and J093012 were combined to maximize the
available observations.  J093010's light curve, folded at
19\,674.574~s, then showed a typical contact binary shape, but with
significant non-Gaussian data scatter below the main curve.  A visual
examination of the object's full light curve suggested the cause was
additional deep eclipses on certain individual nights, implying a
second eclipsing body in the field of view.  A frequency power
spectrum also supported an additional periodic signal near 1.3~d.
Subtracting the median binned light curve (corresponding to the
contact binary) from the data revealed the light curve of an EA-type
(detached) eclipsing binary with period 112\,799.109~s.

The juxtaposition of these two binaries on the sky did not seem to be
coincidental.  Two sources had been observed at this location by
Hipparcos as TYC~3807-759-1 and TYC~3807-759-2, with equivalent
Johnson $V$ magnitudes 9.851 and 10.990, respectively; a separation of
1\farcs88; and a proper motion of pmRA:
-8.0~mas~yr\textsuperscript{-1}, pmDE:
-9.4~mas~yr\textsuperscript{-1}, measured at their joint photocentre.
The tiny angular separation and similar magnitudes favoured a
plausible interpretation of the two eclipsing binary systems as being
gravitationally bound in a quadruple doubly eclipsing system.

Only five other doubly eclipsing quadruple systems had been proposed
at the time of our announcement of this discovery in \citet{lohr13}:
BV~Dra+BW~Dra \citep{batten}, V994~Her \citep{lee08},
OGLE-LMC-ECL-16545 \citep{graczyk}, KIC~4247791 \citep{lehmann}, and
Cze~V343 \citep{cagas}.  The contact binary in J093010 had a shorter
period than any system in the other five quadruples, making it
particularly amenable to further observations.

\citet{koo} obtained BV photometry over several months in
2012--2013 for each eclipsing binary separately, and time-series
spectra for the whole system.  Their light curves
made clear that the brighter signal (TYC~3807-759-1) was associated
with the detached binary, and they consequently termed this source
J093010A, and the fainter source J093010B (containing the contact
binary).  We follow this convention here.  Although \citeauthor{koo}
were unable to detect a spectral signature of the assumed contact
binary in their combined spectra (due in part to the length of their
exposures), they did observe clear line splitting and shifting with
the period of the proposed detached binary, confirming its reality.
They also unexpectedly detected an additional set of static spectral
lines which they interpreted as a third component in J093010A, which
would make the whole system a highly unusual doubly eclipsing
quintuple.

Independently, we obtained time-series spectroscopy over three nights
in 2012--2013 for J093010A and J093010B separately, and
near-simultaneous RGB photometry for the whole system to help
establish the phases of the observations.  Our new data sets thus
complement those provided in \citet{koo}, allowing us to model
both binaries quite fully; the revisited SuperWASP observations also
provide further long baseline information.
  
\subsection{Spectroscopy}

\begin{figure}
\begin{minipage}[t]{0.1\linewidth}
\centering
\includegraphics[width=\textwidth]{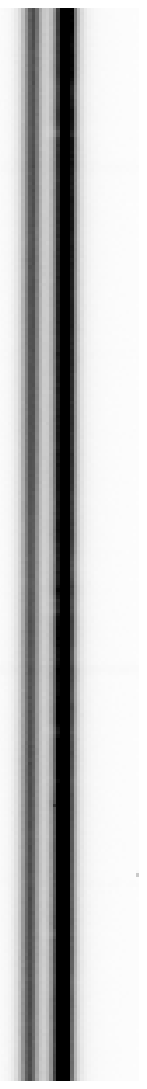}
\end{minipage}
\qquad
\begin{minipage}[t]{0.75\linewidth}
\centering
\includegraphics[width=\textwidth]{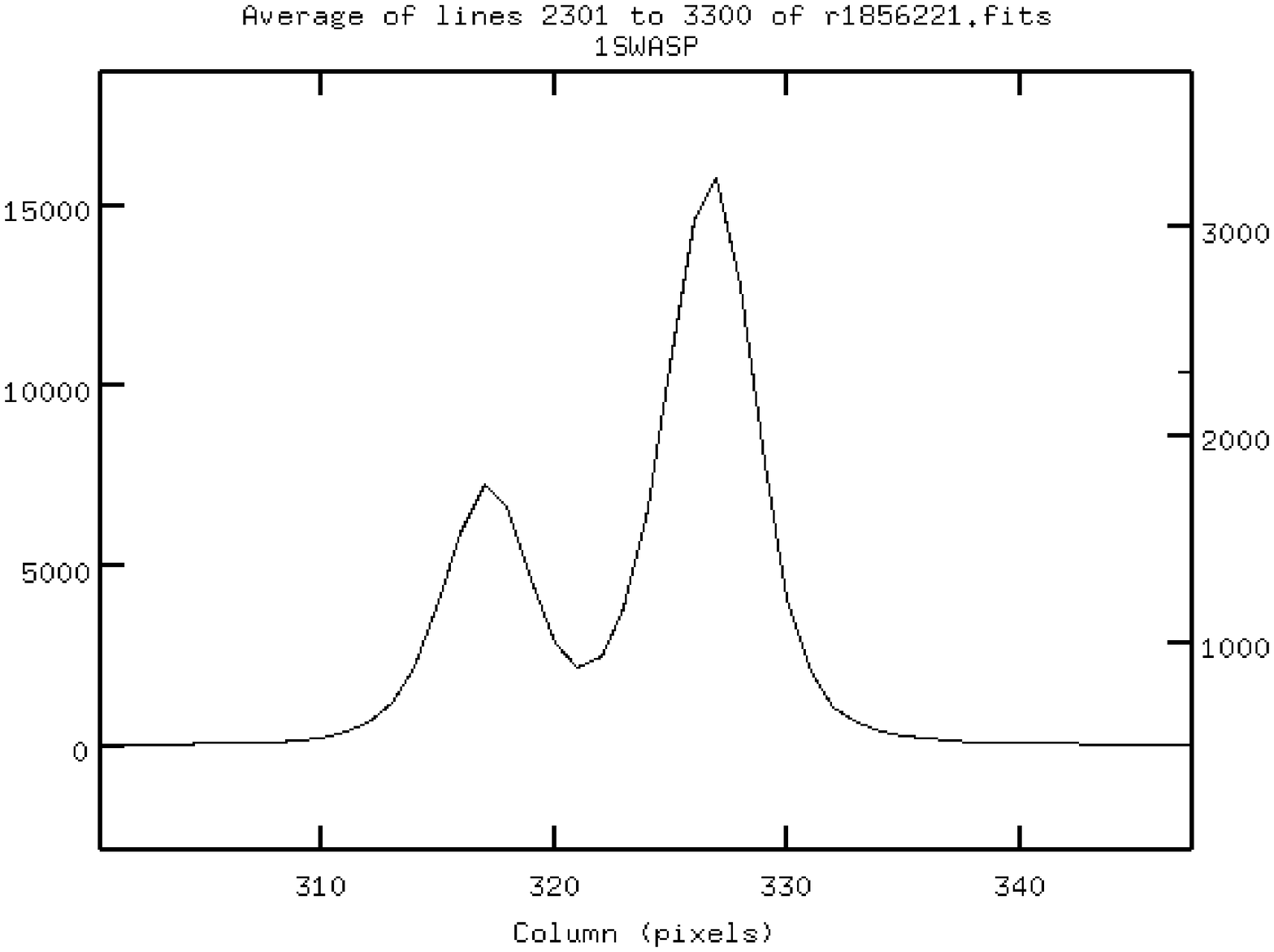}
\end{minipage}
\caption{Left: Section of spectra for J093010A and J093010B from a
  single 360~s exposure, showing spatial separation.  Right: Plot of
  spectral profile from same exposure, cut perpendicular to the
  dispersion axis, showing two clear peaks in the line dispersion
  function.  The stronger peak on the right-hand side corresponds to
  J093010A.}
\label{quintspecplate}
\end{figure}

\addtocounter{table}{1}

49 long-slit spectra were obtained by E.~Gillen for J093010 on 22 and
31 December 2012, and on 2 January 2013, using the red arm of the ISIS
spectrograph on the 4.2~m William Herschel Telescope at La Palma.  The
R1200R grating provided an intermediate resolution of 0.24~\AA\ per
pixel.  A usable wavelength range of $\sim$8350--9000~\AA\ was
obtained, to capture the \ion{Ca}{II} triplet important in low-mass
stars.  Exposure lengths ranged from 5 to 360~s (see
Table~\ref{J093010rvtable}): short enough in all cases to avoid
motion blur for the contact binary.

The images were flat-fielded and bias-corrected by E.~Gillen as part
of a larger observing programme.  On all but one image, two
partially overlapping dispersion lines were clearly visible
(e.g. Fig.~\ref{quintspecplate}), corresponding to J093010A and
J093010B, and we were able to extract separate spectra for each, using
standard IRAF tools and optimal extraction.  Wavelength calibration
was carried out using arc spectra taken with CuArNe lamps.  S/N ratios
reached 340 for the brighter J093010A and 220 for the fainter J093010B
in the longest exposures, when both systems were near quadrature, and
fell to 45 and 25 respectively in the initial 5~s exposure.

Preliminary radial velocities were extracted using the IRAF task
FXCOR, using a synthetic spectrum with $T$=4500~K and log~$g$=4.5 as
the comparison template.  These allowed approximate determination of
mass ratios and radial velocity curve amplitudes, which provided
initial input parameters for spectral disentangling of J093010A and B.
The KOREL code \citep{hadrava}, implemented on the Virtual
Observatory\footnote{https://stelweb.asu.cas.cz/vo-korel}, was used to
disentangle sections of the spectra prepared with
\citeauthor{hadrava}'s PREKOR software, giving improved values for
mass ratio and radial velocities.  The disentangled spectra were also
compared with appropriately broadened template spectra extracted from
the PHOENIX synthetic stellar library
\citep{husser}\footnote{phoenix.astro.physik.uni-goettingen.de}, to
estimate the temperatures of individual components.

\subsection{Photometry}

67 images of J093010 (55 each in Baader G and B filters, 57 in R) were
taken with the Open University's robotic 0.425~m PIRATE telescope
\citep{holmes,kolb14} in Mallorca on the night of 30--31 December
2012, monitored by R.~Busuttil.  All exposures were 80~s; short enough
to avoid phase smearing for the shorter-period system.  Corrections
for bias level, dark current and flat-fielding were made to the frames
using standard IRAF tools, and four comparison stars were identified
on the frames: TYC~3807-1509-1, TYC~3807-54-1, TYC~3807-1503-1 and
TYC~3807-621-1.  These had catalogue colours indicative of classes
between late F and G, and V magnitudes between 9 and 10, similar to
J093010.  They were checked for short and long-term variability with
their SuperWASP light curves, and did not exhibit significant
variations.  Aperture photometry was carried out on all the stars
using the IRAF APPHOT package, and the light curves of the comparison
stars were combined in IDL.  Differential light curves were then
obtained for J093010 relative to this combined comparison curve.

The SuperWASP archive data for J093010 were also reconsidered: there
are 5964 photometric points in V, observed between October 2007 and
April 2009.  (The fainter duplicate object J093012 has 5950 points
covering the same time span; its data were used here as a check on the
results found for J093010.)  For reference, SuperWASP exposure lengths
are 30~s.  Initial values for the orbital periods of the two eclipsing
binaries were found with a custom IDL code described in
\citet{lohr14b}: 19674.594$\pm$0.005~s and 112799.10$\pm$0.15~s; then
a second code was written to separate the two eclipsing signals.  The
light curve was first folded on the shorter period corresponding to
the contact binary, and phase-binned to give a smooth mean curve (90
bins were used); an optimally-weighted average was then found for the
data points in each bin (which corresponded closely to the visible
contact binary signal).  A spline curve was interpolated to these
binned average points, and subtracted from the full light curve.  The
residue was then folded on the longer period corresponding to the
detached binary, and a binned average curve obtained as before.  This
was again subtracted from the full light curve, to leave a clean
contact binary light curve as output.  This clean mean contact curve
was subtracted a second time from the full data set, to leave a clean
detached binary light curve for output.  Our initial period-searching
code was run again on the separated files, and its resulting period
values were used to initialize the signal-separation code for a second
iteration.  This continued until convergence was reached for both
periods.

The BV light curves of \citet{koo} were also used in our modelling;
most details of the photometry are given in that paper, though we add
that the exposure lengths were below 25~s, avoiding any risk of
phase-smearing for even the shorter-period binary.

\section{Results}

\subsection{Orbital periods}

\begin{figure}
\resizebox{\hsize}{!}{\includegraphics{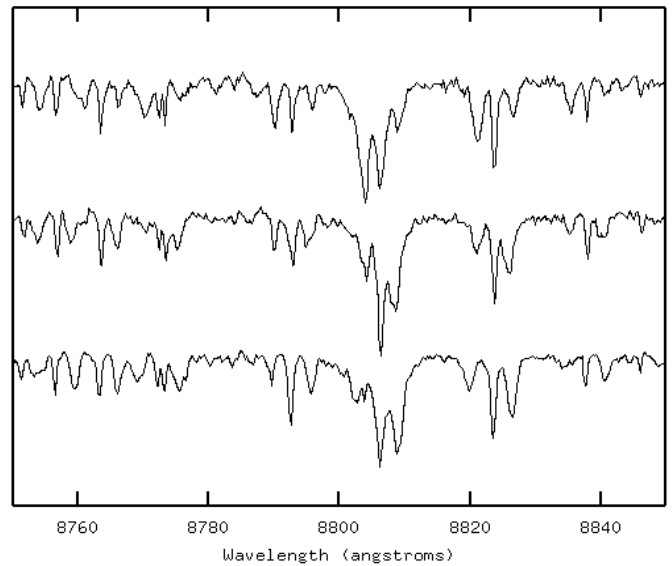}}
\caption{Extract of WHT spectra for J093010A at approximate phases
  0.16, 0.64 and 0.75 (top to bottom).  Relatively isolated absorption
  lines at around 8793, 8806 and 8823~\AA\ exhibit three-way
  splitting, with the outer pair also shifting between phases.}
\label{triplespecs}
\end{figure}

\begin{figure}
\resizebox{\hsize}{!}{\includegraphics{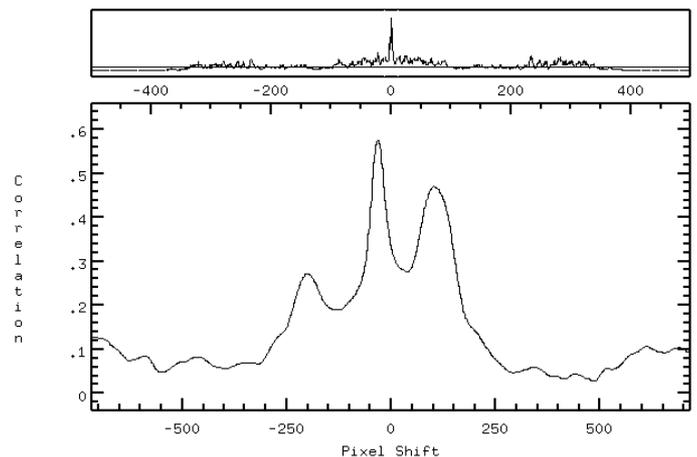}}
\caption{Cross-correlation plot for J093010A at phase 0.75,
  exhibiting three strong peaks corresponding to three stellar
  components (primary on the right, secondary on the left, tertiary in
  the middle).  The broadest absorption lines were excluded from the
  cross-correlation to minimize the radial velocity uncertainties.}
\label{tripleccf}
\end{figure}

The orbital periods based on SuperWASP data rapidly converged on
19674.47$\pm$0.03~s for the contact binary and 112798.90$\pm$0.16~s
for the detached binary i.e. 0.2277138(3)~d and 1.3055428(19)~d
respectively.  (The periods for J093012 did not converge, but
oscillated within the narrow ranges 19674.43--19674.52 and
112798.9--112800.0, which include and support the results for
J093010.)  These periods fall within the uncertainties of those found
by \citeauthor{koo} from their photometry: 0.2277135(16)~d and
1.30550(4)~d, though their figure for the detached binary's period is
based on a single measured primary eclipse time and three secondary
eclipses.

No period change was detected within the SuperWASP data for either
binary; extrapolating forward from the most precisely measured times
of minimum for the SuperWASP data, using our periods, gives an O$-$C
value of just $-$30~s for the \citeauthor{koo} primary minimum for the
detached system, and $\sim$+200~s for their primary minima for the
contact system.  The small size of these discrepancies further
supports both the absence of significant period change (especially in
the detached system) and the reliability of our periods, calculated
over long base lines using whole light curves, rather than minimum
timings alone.  We therefore prefer our periods for the remainder of
the analysis, but use \citeauthor{koo}'s more precise and recent
primary minimum timings to calculate phases.

\subsection{Radial velocities and spectroscopic parameters}

\citeauthor{koo} measured radial velocities for J093010A by fitting
three Gaussians simultaneously to five isolated absorption lines in
their spectra, in the range 6400--6800~\AA.  Our longer wavelength
range contained such a profusion of blended lines that this approach
would have been impossible; indeed, it was hard to detect three-way
line splitting from visual examination of the spectra over most of the
range (Fig.~\ref{triplespecs} illustrates a small region where the
triplets are relatively isolated).  However, cross-correlation
revealed three extremely clear correlation peaks for every spectrum
(e.g. Fig.~\ref{tripleccf}), allowing preliminary radial velocities to
be measured easily for the system.

\begin{figure}
\resizebox{\hsize}{!}{\includegraphics{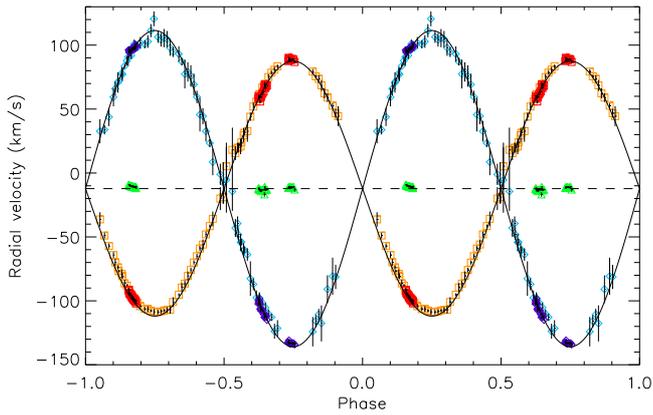}}
\caption{Radial velocity curves for J093010A.  The primary's
  measurements are indicated by squares, the secondary by diamonds and
  the third component by triangles.  Data from \citet{koo} is in
  fainter print (orange and light blue in the online version of the
  paper) and our new results are in bold print (red, dark blue and
  green).  Phase-folding uses \citeauthor{koo}'s primary minimum time
  of HJD~2456346.78443 and our optimum period.  PHOEBE model fits are
  overplotted (solid black curves) and the location of the modelled
  system velocity is shown by the dashed line.}
\label{triplervplot}
\end{figure}

Spectral disentangling was then achieved for three subregions of our
spectra containing narrow and well-defined lines, and converged on
consistent values for mass ratio ($q=0.806\pm0.007$) and
semi-amplitude of the primary's radial velocity curve
($K_1=99.4\pm0.7$~km~s\textsuperscript{-1}).  The resulting radial
velocities were corrected to heliocentric values by disentangling
telluric lines observed in the region 8250--8320~\AA, and are given in
Table~\ref{J093010rvtable}.  (These are very close to those found by the
cross-correlation method, but with smaller uncertainties.)
Fig.~\ref{triplervplot} shows our results, together with those of
\citeauthor{koo}: where our observations overlap with theirs (around
phases 0.15 and 0.65) there is close agreement in both the amplitudes
and absolute values of the curves; we have also fortuitously been able
to observe the system around its secondary maximum (phase 0.75) where
\citeauthor{koo} had a gap.

Also notable is our confirmation of the third strong and near-static
component of the detached system spectra: \citeauthor{koo}'s fifth
star.  Since we have obtained separate spectra for J093010A and
J093010B, we can be certain that this additional source occurs at the
same location on the sky as the detached eclipsing binary, rather than
being near the contact system or mid-way between the two.  There is
also very good reason to regard it as part of a gravitationally bound
system including the detached binary: its radial velocities (average
$-11.6\pm1.5$~km~s\textsuperscript{-1}) are visibly very close to the
cross-over system velocity of the binary (modelled as
$-12.3\pm0.2$~km~s\textsuperscript{-1}).  While longer-term radial
velocity and astrometry measurements would be required to demonstrate
convincingly that the fifth star forms a triple subsystem with the
detached binary (rather than, for example, the two binaries being more
closely bound to each other), the coincidence of its spectrum with
that of the binary in J093010A (see Fig.~\ref{quintspecplate}) makes
this a more plausible hierarchy.

\begin{figure}
\resizebox{\hsize}{!}{\includegraphics{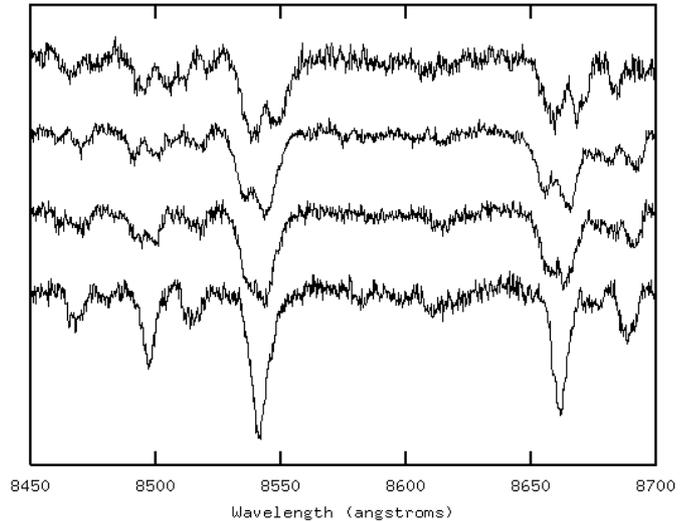}}
\caption{Extract of WHT spectra for J093010B at approximate phases
  0.24, 0.66, 0.88 and 0.99 (top to bottom).  The \ion{Ca}{II} triplet
  of absorption lines at around 8498, 8542 and 8662~\AA\ exhibit clear
  splitting and shifting between phases.}
\label{contactspecs}
\end{figure}

\begin{figure}
\resizebox{\hsize}{!}{\includegraphics{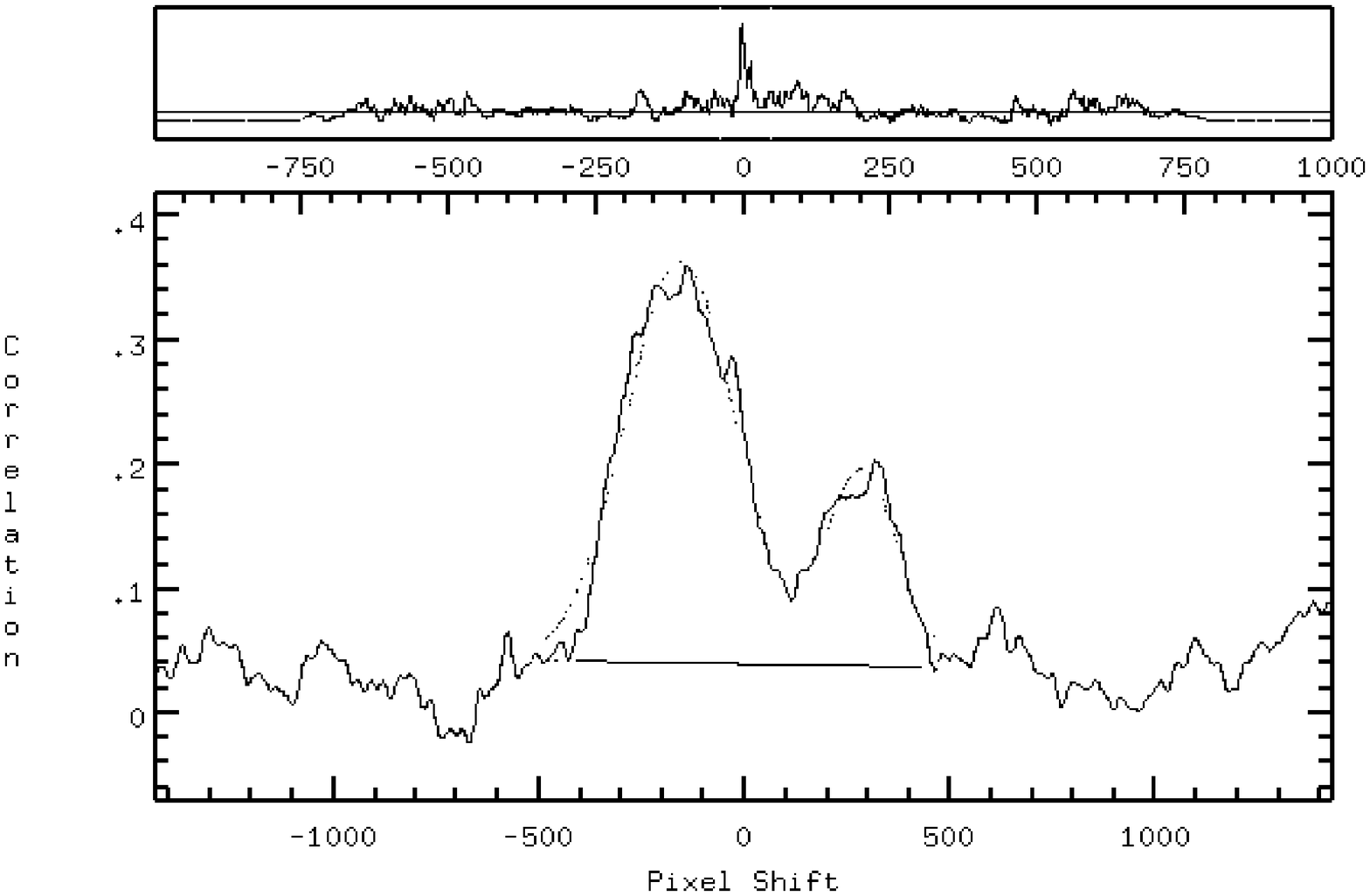}}
\caption{Cross-correlation plot for J093010B at phase 0.24,
  exhibiting two strong peaks corresponding to the binary components
  (primary on the left, secondary on the right).  The two broadest
  calcium lines at 8542 and 8662~\AA\ were excluded from the
  cross-correlation to minimize the uncertainty in the radial velocity
  measurements.}
\label{contactccf}
\end{figure}

We were also able to measure radial velocities from most of the
spectra for the candidate contact system in J093010B.  Here, splitting
and shifting of the strongest lines (primarily the \ion{Ca}{II}
triplet) was apparent from visual inspection of the spectra alone
(e.g. Fig.~\ref{contactspecs}), confirming this system as a
double-lined spectroscopic binary also.  The results from
cross-correlation (e.g. Fig.~\ref{contactccf}) again provided starting
values for spectral disentangling, and here we were able to
disentangle the full spectrum as a single region, giving much smoother
radial velocity curves with much smaller uncertainties than had been
achieved by cross-correlation (Table~\ref{J093010rvtable}).  A mass ratio of
$0.397\pm0.006$ and $K_1=105.1\pm1.3$~km~s\textsuperscript{-1} were
converged upon by disentangling; a slightly different value of
reference minimum was also implied by the spectra: HJD~2456288.87687,
where \citeauthor{koo}'s nearest minimum (based on photometry) was
HJD~2456288.87879.

\begin{figure}
\resizebox{\hsize}{!}{\includegraphics{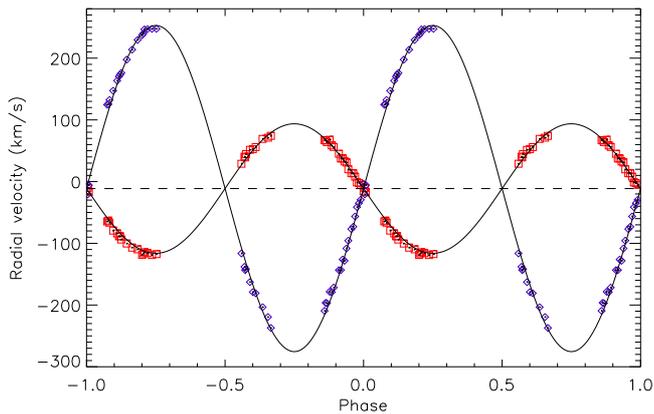}}
\caption{Radial velocity curves for J093010B, based on our WHT
  observations.  The primary is shown with red squares and the
  secondary with blue diamonds; uncertainties are generally smaller
  than the symbol size.  Phase-folding uses HJD~2456288.8780, found
  by PHOEBE modelling of radial velocity and light curves
  simultaneously, and our optimum period.  The dashed line shows the
  location of the (modelled) system velocity.}
\label{binaryrvplot}
\end{figure}

Our results are shown in Fig.~\ref{binaryrvplot}: we have managed to
capture the system around its primary maximum (phase 0.25) and on each
side of its secondary maximum (near phases 0.6 and 0.9), giving
sufficient phase coverage for decent modelling (though further
observations around phases 0.4 and 0.75 would still be desirable).
The cross-over velocity of the two curves is
$-11.3\pm0.7$~km~s\textsuperscript{-1}: very close to the detached
binary system velocity and the third component velocity in J093010A,
strongly supporting the hypothesis that the two binaries and the fifth
star are gravitationally bound within a quintuple system with a common
motion relative to the Sun.

\begin{figure}
\resizebox{\hsize}{!}{\includegraphics{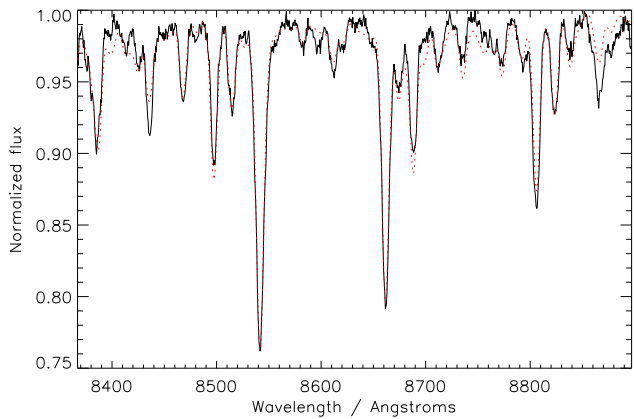}}
\caption{Disentangled renormalized spectrum for primary component of
  J093010B (black solid lines), with best-fitting PHOENIX synthetic
  template overplotted (red dotted lines).  The prominent absorption
  lines are \ion{Ca}{II}, \ion{Mg}{I}, \ion{Fe}{I}, \ion{Ti}{I} and
  \ion{Al}{I}.}
\label{contprimspecfit}
\end{figure}

\begin{figure}
\resizebox{\hsize}{!}{\includegraphics{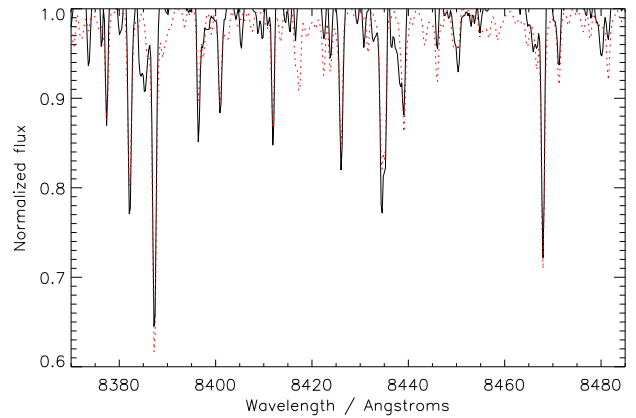}}
\caption{Part of disentangled renormalized spectrum for tertiary
  component of J093010A (see caption to Fig.~\ref{contprimspecfit}).}
\label{terspecfit}
\end{figure}

The disentangled spectrum for the primary component of J093010B was
best fitted by a PHOENIX synthetic template with $T=4700\pm50$~K
and log $g$=4.5 (Fig.~\ref{contprimspecfit}).  The secondary
component's spectrum was of very similar shape, but had far lower S/N
and so was not used to assess the common temperature of the contact
binary.  The temperature found here spectroscopically is entirely
consistent with that suggested by \citeauthor{koo}'s $B-V$ colour for
J093010B during secondary eclipse i.e. $4680\pm50$~K.

The three components of J093010A were difficult to disentangle
convincingly owing to heavy blending of strong lines and our rather
limited phase coverage; however, a reasonably good fit to the tertiary
component was achieved, since it exhibited narrower, deeper lines than
the detached binary components, presumably owing to greater rotational
broadening in the latter.  Templates with $T=5100\pm200$~K and log
$g$=4.5 (Fig.~\ref{terspecfit}) provided the best matches, suggesting
an early K spectral type for this star.  Measurements of the
equivalent widths of several distinct triplets of lines in the
original spectra also indicated that the primary's lines were somewhat
stronger than those of the tertiary, which could imply a slightly
higher temperature (and mass).  Spectroscopy with better phase
coverage would be desirable in future to allow direct determination of
the temperatures of all three components in J093010A.  However, we
were able to estimate temperatures for the detached binary components
from the photometry, during the modelling process described in
Sect.~\ref{sect_a}.

\subsection{Light curve characteristics}

\begin{figure}
\resizebox{\hsize}{!}{\includegraphics{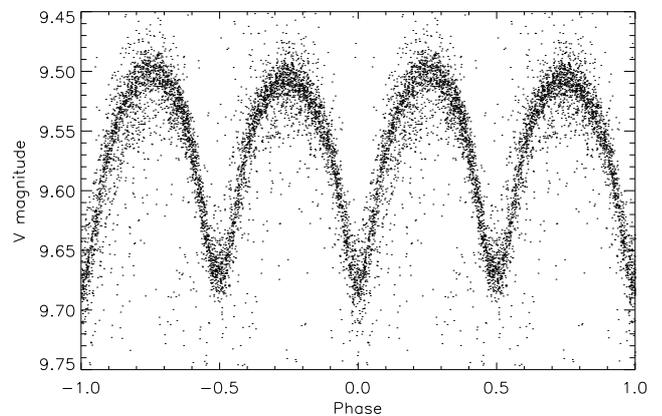}}
\caption{SuperWASP folded light curve for J093010B, converted to V
  magnitudes.  The maximum magnitude is that of the quintuple system
  as a whole.}
\label{swbinarylcplot}
\end{figure}

Moving to the photometric results, we can compare the light curves
obtained by \citet{koo} for J093010A and B separately with the
SuperWASP light curves extracted from the combined light variation for
the whole system.  In each case, the SuperWASP amplitudes are smaller,
and the magnitudes lower, because there is an effective third light
included containing the (maximum) contributions from the other three
stars.  For the contact binary the SuperWASP curve
(Fig.~\ref{swbinarylcplot}) is evidently the same general shape as the
BV curves (Fig.~2 in \citeauthor{koo}): the secondary eclipses are
visibly flat-bottomed with the primary minimum fractionally deeper
than the secondary; the secondary maximum is also slightly lower than
that of the primary, though this effect is perhaps stronger in
\citeauthor{koo}'s curves, which would support the presence of a
long-term static spot on one of the components.

\begin{figure}
\resizebox{\hsize}{!}{\includegraphics{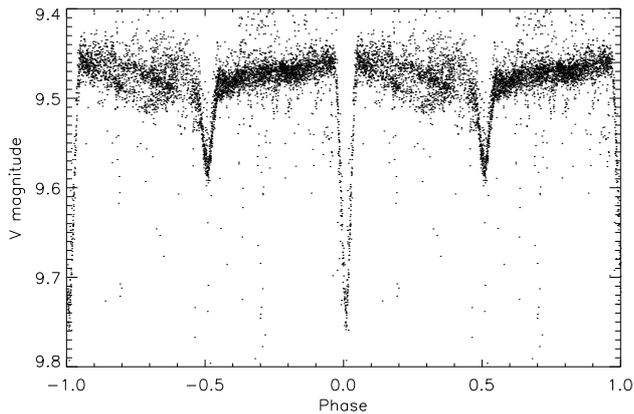}}
\caption{SuperWASP folded light curve for the detached binary in
  J093010A, converted to V magnitudes.  The maximum magnitude is that
  of the quintuple system as a whole.}
\label{swtriplelcplot}
\end{figure}

\begin{figure}
\resizebox{\hsize}{!}{\includegraphics{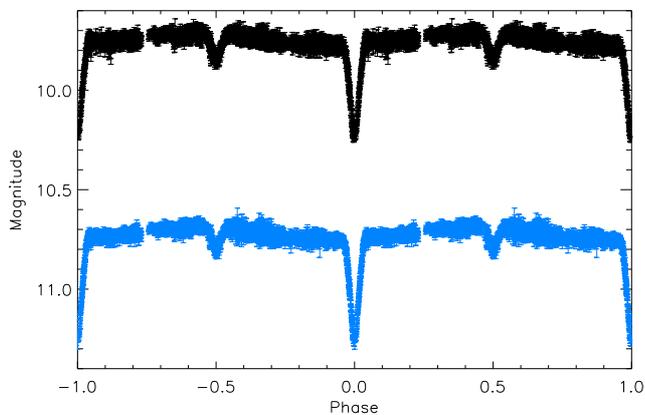}}
\caption{\citeauthor{koo} light curves for the detached binary in
  J093010A, in V (top) and B (bottom), converted from differential
  instrumental magnitudes to standard magnitudes using normalization
  indices provided in \citeauthor{koo} (their Table~6).}
\label{kortriplelcplot}
\end{figure}

For the detached binary light curves in J093010A
(Figs.~\ref{swtriplelcplot} and \ref{kortriplelcplot}), the
relative depths of the two eclipses are comparable in the SuperWASP
data and \citeauthor{koo}'s data, and together with the sharp
ingresses and egresses of the eclipses they support a well-detached
EA-type system.  The eclipses are equally spaced and of the same
duration, indicating no significant orbital eccentricity.  The primary
eclipse appears flat-bottomed while the secondary eclipse's shape is
arguably more curved; if taken at face value this would suggest that
the secondary's radius is (surprisingly) greater than that of the
primary.  The out-of-eclipse region would be expected to be
essentially flat in an EA-type light curve, but is not.  In the
SuperWASP curve, the region between phases 0.1 and 0.4 is far more
scattered and noisy than the region between phases 0.6 and 0.9;
moreover, the maximum is near the primary minimum, while it occurs
just before the secondary minimum in \citeauthor{koo}'s curves; they
suggest that this latter feature results from a moving spot on one of
the components.

To explore this idea further, we split up the SuperWASP detached
binary curve by time, into three sections of contiguous data covering
several months.  The profile of the out-of-eclipse region clearly
changed over time, with its minimum occurring around phase 0.8 near
HJD~2454510, phase 0.4 near HJD~2454820 and phase 0.3 near HJD~2454870
i.e. apparently moving negatively in phase at a rate of (very) roughly
one cycle every two years.  The superposition of these three (fairly
smooth) curves produced the apparent scatter in the curve of
Fig.~\ref{swtriplelcplot}.  The most plausible explanation for this
moving profile would indeed seem to be a spot on the surface of one of
the detached system components, rotating at a slightly different rate
from the star as a whole, perhaps because of differential rotation.

\begin{figure}
\resizebox{\hsize}{!}{\includegraphics{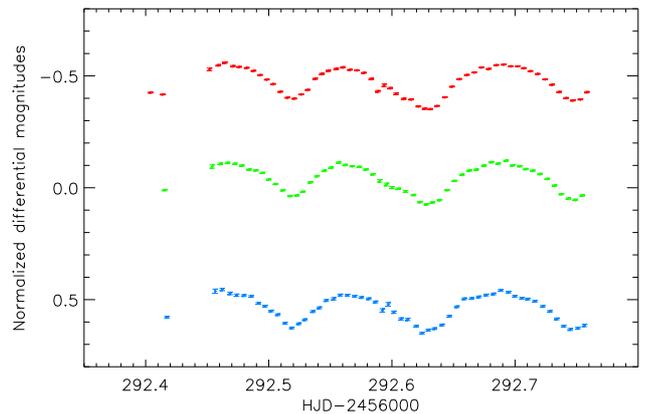}}
\caption{Normalized differential PIRATE light curves for J093010 in Baader R
  (red, top), G (green, middle) and B (blue, bottom) filters.}
\label{quintdiffmags}
\end{figure}

Finally, Fig.~\ref{quintdiffmags} shows the PIRATE RGB photometry
obtained for the whole system during one night.  As predicted by the
SuperWASP ephemeris, contact binary primary minima occur around HJD
2456292.519 and 2456292.748, while the two eclipsing systems'
secondary minima overlap in between (this is most clearly seen in the
G curve, where the detached system secondary minimum occurs around HJD
2456292.60 and the contact one around HJD 2456292.63).  The second
night of our WHT spectra begins just as this photometry ends, and
covers phases $\sim$0.07--0.25 of the contact system, which is
entirely consistent with the light curve.  Additional nights of
photometry would be useful to establish the average shapes of the
three-colour light curves; however, we note that the three curves have
very similar shapes.  The primary eclipses of J093010B are perhaps
fractionally deeper in B than R, with G intermediate.  Similarly,
\citeauthor{koo}'s standardized BV magnitudes for J093010A indicate
that the primary eclipses are deeper in B, while the secondary
eclipses are deeper in V i.e. the primary is hotter than the average
of J093010A's three components, while the secondary is cooler.

\subsection{Flux contributions}

We can make further estimates of the relative flux contributions of
the five stars from the photometry and spectroscopy.  Starting with
the SuperWASP light curves in SuperWASP flux units, if we assume that
J093010A and J093010B's angles of inclination are close to
90$^{\circ}$ (since both systems contain flat-bottomed eclipses
supporting near-totality), the depth of each component's eclipse gives
a minimum value for its flux contribution (J093010A primary: 38,
secondary: 15; J093010B primary: 25, secondary: 23).  Adding these
four fluxes and subtracting them from the total flux (160) when no
component is eclipsed gives us a maximum flux estimate for the fifth
component (59).  These values would suggest that the tertiary makes up
at most 53\% of the flux in J093010A (in V), and the primary and
secondary at least 34\% and 13\% respectively.  \citeauthor{koo}'s V
band data similarly supports contributions of around 53\%, 37\% and
10\% respectively, using the same approach; in B the proportions are
53\%, 40\% and 8\%.

Furthermore, the SuperWASP fluxes allow us to estimate
J093010B:J093010A flux contributions at the former's maximum and
primary minimum as 43\% and 21\%.  These figures can be compared with
the spectroscopic fluxes, found from fits to the continua of the
extracted spectra, evaluated at 8500~\AA\ and scaled to account for
the different exposure lengths: near the maxima of both systems (phase
0.75 of J093010A and 0.63 of J093010B) the ratio is 44\%, and near
contact primary minimum (around phase 0.18 of J093010A and 0.0 of
J093010B) it is 24\%, showing considerable similarity across optical
and near-infrared observations.  These consistent estimates provide a
starting point for further modelling of the five stars.

\section{Modelling and Discussion}
\label{sect_a}

To determine parameters for the eclipsing components of J093010, the
two binaries were modelled separately using the PHOEBE software
\citep{prsa}, built upon the code of \citet{wildev}.

The starting assumptions for J093010A, based on its light curve, were
that it is a detached binary in a circular orbit viewed nearly
edge-on, with $M_1 > M_2$, $T_1 > T_2$, a third light
contributing up to half the total flux, and a spot on one component.
The mass ratio ($q=0.806$, where \citet{koo} had found 0.841 by
modelling) found from spectral disentangling was set as a
non-adjustable parameter; the period was also fixed at 1.3055428~d
(from the SuperWASP results) and the reference $HJD_0$ at
2456346.78443 (from \citeauthor{koo}'s single primary minimum timing);
data was not binned or converted to phases prior to modelling.  With
the angle of inclination $i$ set to 90\degr, the radial velocity
curves were first fitted for $a \sin i$ (a proxy to the semimajor
axis) and system velocity $\gamma_0$ simultaneously, and the optimum
values found were not adjusted subsequently, since light curves are
largely insensitive to these parameters.

Fig.~\ref{triplervplot} shows the best fit obtained for the
combination of our radial velocity data and that of \citeauthor{koo}.
Similar results were found when modelling our data sets separately: $a
\sin i$ of $5.765\pm0.014$~R$_{\sun}$ and $5.607\pm0.014$~R$_{\sun}$
for us and them respectively; and $\gamma_0$ of
$-12.3\pm0.2$~km~s\textsuperscript{-1} and
$-10.4\pm0.2$~km~s\textsuperscript{-1}.  Combining the data sets
reduced the formal uncertainties on the two parameters in comparison
with each data set considered alone ($a \sin
i=5.759\pm0.010$~R$_{\sun}$ and
$\gamma_0=-12.24\pm0.17$~km~s\textsuperscript{-1}), and this
combination of radial velocity data was used for the remainder of the
modelling.

\begin{figure}
\resizebox{\hsize}{!}{\includegraphics{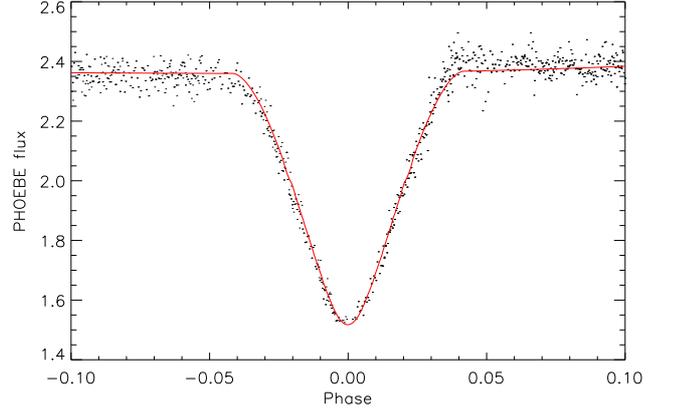}}
\caption{PHOEBE best model fit (red solid curve) to
  primary eclipse of J093010A in V, showing good match to eclipse
  width, shape and depth.}
\label{triplevfitcloseup}
\end{figure}

To model \citeauthor{koo}'s light curves, a valid third light
contribution needed to be included, more precise than the $<53$\%
estimated above.  While \citeauthor{koo} seem to have fitted their
curves treating third light as a free parameter, we feared there might
be strong correlations between this and other fitted parameters, and
so sought to constrain the value of the third light independently.
Since the eclipses of J093010A appear total-annular, we could
determine the phases of first, second, third and fourth contact
($\phi_{1,2,3,4}$) quite precisely for the primary eclipse (see
Fig.~\ref{triplevfitcloseup}), using \citeauthor{koo}'s BV photometry,
and then estimate the radii using these formulae from
\citet{hilditch}:
\begin{eqnarray}(\phi_2-\phi_1)=(\phi_4-\phi_3)&=&\frac{2R_2}{2\pi a}\\
(\phi_3-\phi_1)=(\phi_4-\phi_2)&=&\frac{2R_1}{2\pi a}.\end{eqnarray}

On the assumption that the primary had the larger radius, this gave
$R_1=0.832\pm0.018$R$_{\sun}$ and $R_2=0.669\pm0.018$R$_{\sun}$ (where
\citeauthor{koo} found $R_1=0.757\pm0.008$R$_{\sun}$ and
$R_2=0.743\pm0.010$R$_{\sun}$).  Then, we made use of an eclipse
modelling guideline from \citet{wilson}: ``the ratio of the depth of
the annular eclipse to the light remaining in the total eclipse is
approximately the square of the ratio of smaller to larger star
radii''.  We measured the depth of the annular (primary) eclipse in V
and B (in PHOEBE flux units) from the light curves, and combined this
with the ratio $(\frac{R_2}{R_1})^2$ to estimate the flux which would
remain in the total (secondary) eclipse, if there were no third light
contribution: this would be the true contribution of the primary star
($1.33\pm0.09$ units in V and $0.56\pm0.04$ units in B).  The
secondary's flux was given directly -- and hence with greater
precision -- as the depth of the total eclipse ($0.275\pm0.011$ units
in V and $0.090\pm0.007$ in B).  By subtracting the primary and
secondary fluxes from the maximum flux in the light curves, the
tertiary's contribution could then be estimated more precisely
($0.90\pm0.09$ units in V and $0.37\pm0.04$ in B) as $36\%\pm4$:
somewhat larger than the fitted values found by \citeauthor{koo} of
31\% in V and 32\% in B.  These values for third light were then fixed
for the rest of the modelling.

The Kopal potentials $\Omega_{1,2}$ for the two eclipsing stars were
then set to values which would reproduce the radii determined above
(7.74 and 8.04 respectively), indicating strongly negative filling
factors, as expected for a well-detached system.  These values gave
excellent matches to the widths of the eclipses
(e.g. Fig.~\ref{triplevfitcloseup}).  The effective temperatures of
both components were also estimated from their $B-V$ colours, using
the individual fluxes found above: these were $4970_{-210}^{+240}$K
for the primary and $4470\pm160$K for the better-determined secondary
(by the same method, the third component was estimated to have
$T_3=4900_{-310}^{+370}$, consistent with the value found from its
disentangled spectrum earlier).  Using these values for $T_1$ and
$T_2$ gave a model which had too shallow a primary eclipse and too
deep a secondary eclipse; their ratio was therefore increased by
increments until both eclipses were equally well reproduced: this
occurred at $T_1=5185$K and $T_2=4325$K.  The angle of inclination $i$
also required a slight decrease from {90\degr} to match the depths of
both eclipses simultaneously (hence $a$ and $\Omega_{1,2}$ were also
fractionally adjusted).  As a further check on these temperatures, we
may note that the ratio of primary to secondary eclipse depths matched
the corresponding ratio of the stars' surface brightnesses, as would
be expected \citep{wilson}.

As noted earlier, the apparently flat-bottomed primary eclipse could
alternatively suggest $R_2>R_1$, and so this possibility was also
explored in the modelling.  Using Equations (1) and (2) with the roles
of $R_1$ and $R_2$ reversed, the components' radii, third light
contribution, $\Omega_{1,2}$ and temperatures were again estimated as
described above; however, no appropriate ratio of temperatures could
be found which would fit both eclipse depths unless the uncertainties
on $T_{1,2}$ were exceeded.  Moreover, the best fit produced an
inconsistency between the eclipse depth ratio and the surface
brightness ratio.  In view of these problems, and the mass-radius
mismatch (which would be challenging to explain in a well-detached
system), we prefer the original model for the remainder of the
analysis.  However, improved photometry, especially of the primary
eclipse, would be desirable in future to confirm it more fully.
 
\begin{figure}
\resizebox{\hsize}{!}{\includegraphics{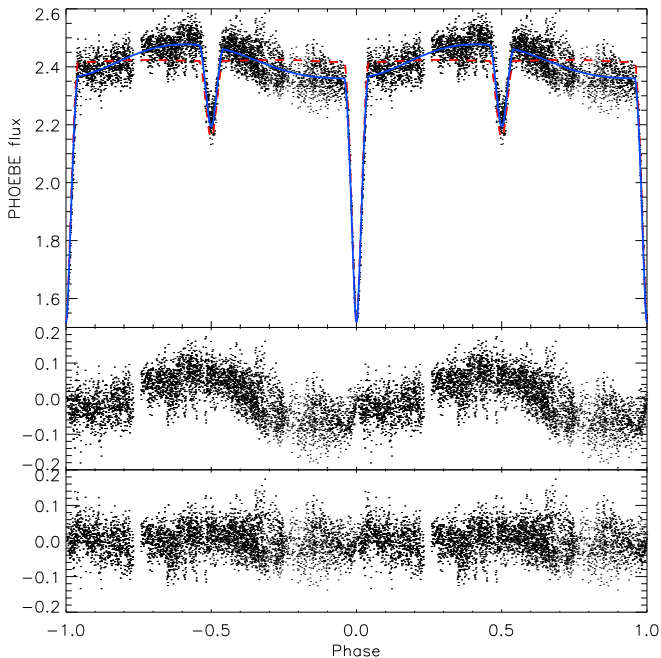}}
\caption{Top: V band light curves for J093010A (data from \citet{koo})
  expressed in PHOEBE flux units, with best-fitting PHOEBE model
  curves overplotted (dashed red lines indicate model with no spot,
  solid blue lines indicate one-spot example model). Middle:
  residuals from no-spot model, showing out-of-eclipse semi-sinusoidal
  light variation unaccounted for.  Bottom: residuals from one-spot
  model, largely correcting for these variations.}
\label{triplevfit}
\end{figure}

\begin{figure}
\resizebox{\hsize}{!}{\includegraphics{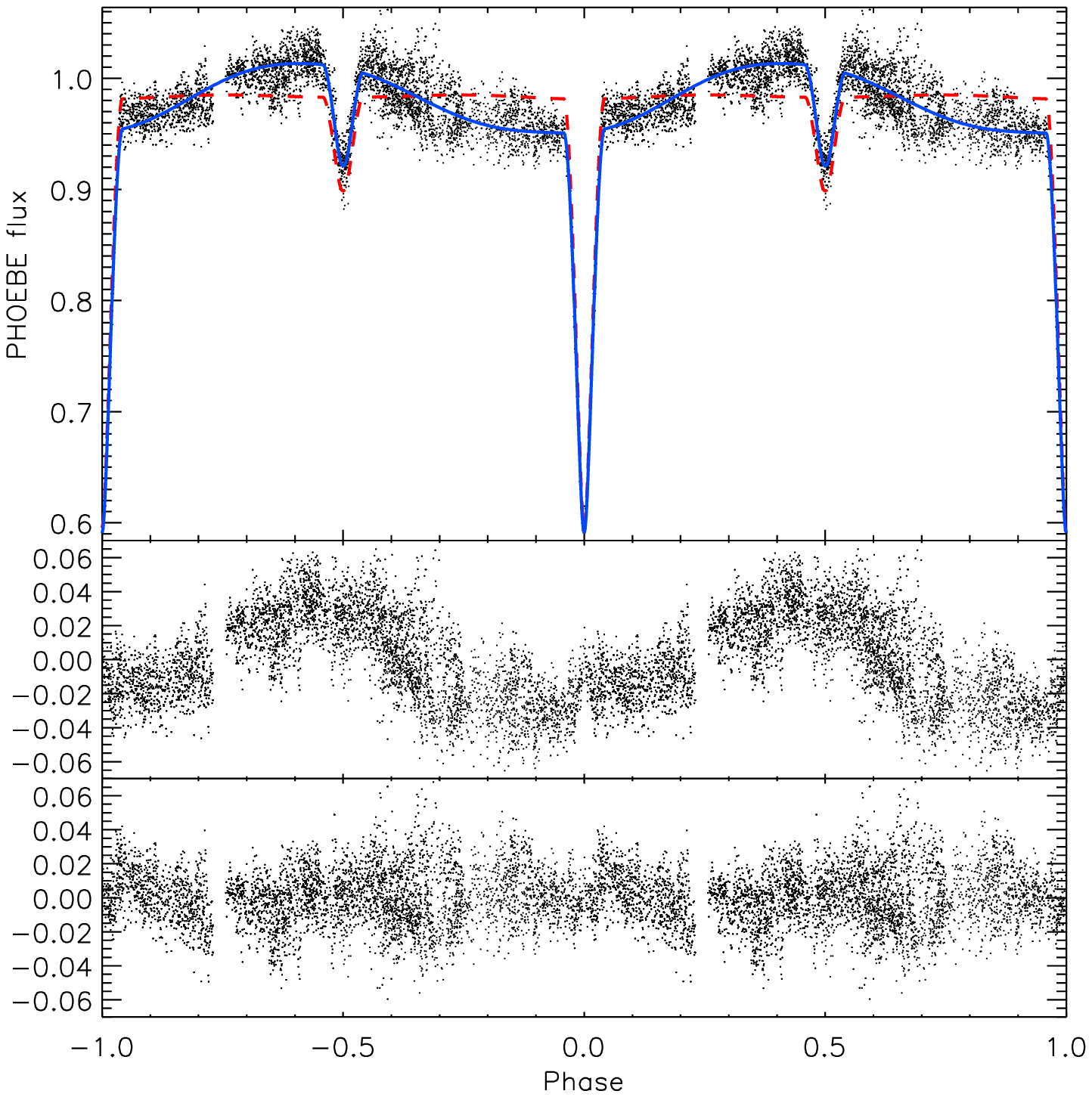}}
\caption{B band light curves, fits and residuals for no-spot and
  one-spot models for J093010A (see caption to
  Fig.~\ref{triplevfit}).}
\label{triplebfit}
\end{figure}

\begin{figure}
\resizebox{\hsize}{!}{\includegraphics{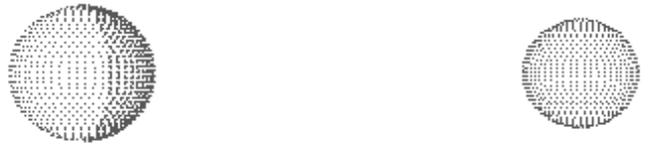}}
\caption{Image of best-fit PHOEBE one-spot model for J093010A at phase
  0.25.  The large cool spot is visible on one limb of the (smaller)
  primary at quadrature.}
\label{detbin}
\end{figure}

The full optimal light curve fits and their residuals are shown in
Figs.~\ref{triplevfit} and \ref{triplebfit}.  It is apparent that,
while the eclipses are very well-fitted by this model, the
out-of-eclipse portions are poorly fitted (reduced $\chi^2=17.1$ in V,
2.62 in B).  This is almost certainly due to the presence of the
assumed moving spot discussed earlier.  To test this explanation, a
single large cool spot was placed on the primary and its four
parameters varied manually until a fair match to the light curve was
achieved (colatitude 90\degr, longitude 29\degr, radius 95\degr and
temperature 0.9835 $\times$ mean primary temperature).  The resulting
light curve fits and residuals are also shown in
Figs.~\ref{triplevfit} and \ref{triplebfit}, and an image of the model
system in Fig.~\ref{detbin}.  These models reproduce the light
variations much better ($\chi^2=8.86$ in V, 1.18 in B).

It should be emphasized that we do not regard this as the only
possible spot configuration which would reproduce the observed light
curve variations: a cool spot on the opposite side of the secondary
also works passably well, and hot spots on either component would also
achieve similar results.  \citeauthor{koo} used two cool spots with
quite different parameters from ours, one on each component in their
preferred model, to achieve a very close fit to the light curve;
however, by appropriate placement of multiple spots, any light curve
features whatsoever can be reproduced, and we did not wish to include
more components in our model than were justifiable.  That said, it is
very likely that there are multiple spots on this system, or
non-circular spots, whose modelling would produce an even better fit
to the light curve, but a method such as Doppler tomography would be
required to determine their nature more rigorously.

Table~\ref{tripleparamtable} gives the final parameters determined for
J093010A, using the no-spot model (there is almost no change to any of
these under the one-spot model).  The uncertainties on $a$, $\gamma_0$
and $i$ are formal errors generated by PHOEBE; the uncertainties on
$q$, $T_{1,2}$, $R_{1,2}$ and $\Omega_{1,2}$ were obtained from
independent measurements as described above; and the uncertainties on
$M_{1,2}$ were found using the formula prescribed in the PHOEBE
manual:
\begin{eqnarray}\sigma_{M_1}&=&M_1\left(3\frac{\sigma_a}{a}+2\frac{\sigma_P}{P}+\frac{\sigma_q}{q+1}\right)\\
\sigma_{M_2}&=&M_2\left(3\frac{\sigma_a}{a}+2\frac{\sigma_P}{P}+\frac{\sigma_q}{q(q+1)}\right).\end{eqnarray}

\begin{table}
\caption{System and stellar component parameters for J093010A}
\label{tripleparamtable}
\centering
\begin{tabular}{l | l l}
\hline\hline
 & Primary & Secondary \\
\hline
Orbital period (s) & \multicolumn{2}{|c}{$112798.90\pm0.16$} \\
Semi-major axis (R$_{\sun}$) & \multicolumn{2}{|c}{$5.762\pm0.010$} \\
Mass ratio & \multicolumn{2}{|c}{$0.806\pm0.007$} \\
System velocity (km~s\textsuperscript{-1}) & \multicolumn{2}{|c}{$-12.24\pm0.17$} \\
Angle of inclination (\degr) & \multicolumn{2}{|c}{$88.2\pm0.3$} \\
Kopal potential & $7.74\pm0.15$ & $8.04\pm0.18$ \\
Filling factor & $-9.4$ & $-10.0$ \\
Mass (M$_{\sun}$) & $0.837\pm0.008$ & $0.674\pm0.007$ \\
Radius (R$_{\sun}$) & $0.832\pm0.018$ & $0.669\pm0.018$ \\
Temperature (K) & $5185_{-20}^{+25}$ & $4325_{-15}^{+20}$ \\
Surface gravity & 4.52 & 4.62 \\
Bolometric luminosity & $5.66\pm0.02$ & $6.92\pm0.02$ \\
Flux contribution (V) & $0.532\pm0.036$ & $0.110\pm0.004$ \\
Flux contribution (B) & $0.552\pm0.040$ & $0.089\pm0.007$ \\
\hline
\end{tabular}
\end{table}

J093010B was simpler to model.  We started with the assumptions that
it is a contact system (i.e. $T_1=T_2=4700$K, $\Omega_1=\Omega_2$) in
a circular orbit seen nearly edge-on, with $M_1 > M_2$, $R_1 > R_2$
and no third light.  The period was set to 0.2277138~d, the mass ratio
to $0.397\pm0.006$, and the reference $HJD_0$ to 2456288.8780 (an
average of the preferred values for radial velocity curves and light
curves, since the continuous variation of J093010B's light curves made
estimation of the time of primary minimum less reliable from
photometry alone).  We first fitted the radial velocity curves alone
for $a \sin i$ and $\gamma_0$, as we did for J093010A, and found the
best fit shown in Fig.~\ref{binaryrvplot}, which gave $a \sin
i=1.661\pm0.009$R$_{\sun}$ and
$\gamma_0=-11.3\pm0.7$~km~s\textsuperscript{-1}.  Here, we note that
\citet{koo} also advanced a model for J093010B based purely on its
light curve, which used a value for the mass ratio of $0.468\pm0.005$
found via the ``q-search'' method: some distance from the value found
here.

\begin{figure}
\resizebox{\hsize}{!}{\includegraphics{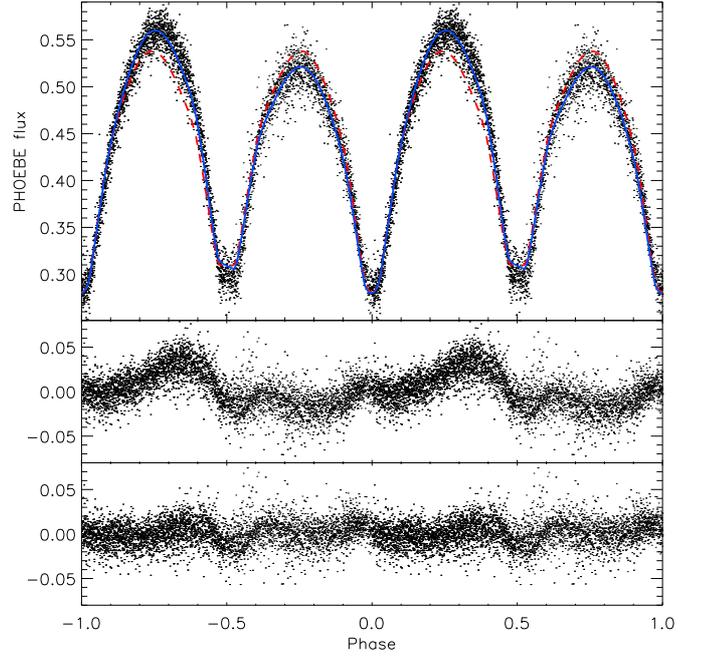}}
\caption{V band light curves, fits and residuals for no-spot and
  one-spot models for J093010B (see caption to
  Fig.~\ref{triplevfit}).}
\label{doublevfit}
\end{figure}

\begin{figure}
\resizebox{\hsize}{!}{\includegraphics{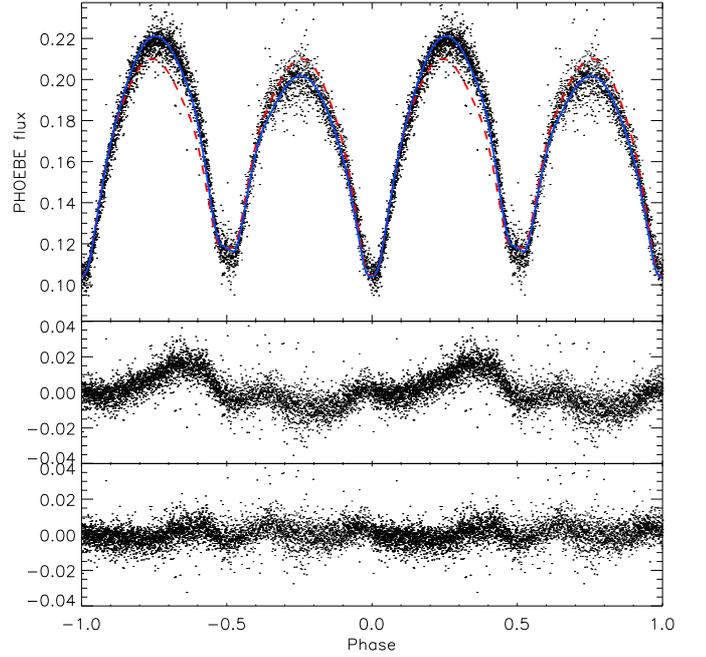}}
\caption{B band light curves, fits and residuals for no-spot and
  one-spot models for J093010B (see caption to
  Fig.~\ref{triplevfit}).}
\label{doublebfit}
\end{figure}

Then, using \citeauthor{koo}'s BV light curves, and again adjusting
$a$ manually to keep $a \sin i$ at the value found earlier, $i$ and
$\Omega$ were allowed to reach their optima simultaneously (the
continuous variation of the light curve meant that the approach taken
to find J093010A's radii was impossible here).  The best fits are
shown in Figs.~\ref{doublevfit} and \ref{doublebfit}.  The Kopal
potential corresponding to this fit ($2.63\pm0.08$) indicates a
positive filling factor of 0.19 as expected for a contact system.  The
angle of inclination found is $86\pm4$\degr, which we note is
consistent with that found for J093010A, suggesting that the two
binaries may share an orbital plane and perhaps originally fragmented
out of a single protostellar disk.  However, measurements of the
longitude of the ascending node for both binaries would be required to
prove a common plane.

\begin{figure}
\resizebox{\hsize}{!}{\includegraphics{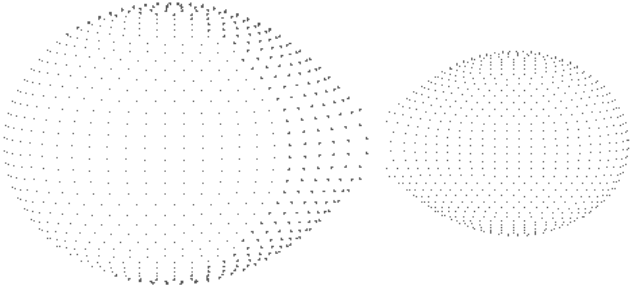}}
\caption{Image of best-fit PHOEBE one-spot model for J093010B at phase
  0.25.  The large cool spot is visible at the limb of the (larger) primary.}
\label{contbin}
\end{figure}

Again, a better fit can be achieved if a spot is modelled, to account
for the different heights of the maxima.  An example spotted model is
shown in Figs.~\ref{doublevfit} to \ref{contbin}; this uses a spot on
the primary with colatitude 90\degr, longitude 60\degr, radius
110\degr and temperature 0.979 $\times$ mean primary temperature.  As
with J093010A, we emphasize that this is only one of several possible
ways to model the light variations using a single spot;
\citeauthor{koo} used a cool spot on the primary and a hot spot on the
secondary instead.  Our resulting parameters for the system, shown in
Table~\ref{doubleparamtable}, are not affected by the presence of the
spot.  The uncertainties on $a$, $q$, $\gamma_0$, $i$ and $M_{1,2}$
were found as for J093010A; that for $T$ was estimated from the
fitting to the disentangled primary spectrum; that for $\Omega$ is a
formal output of PHOEBE; and those for $R_{1,2}$ were found by setting
$a$, $q$ and $\Omega$ to their extrema in appropriate combinations,
which may be expected to overestimate these errors somewhat.

\begin{table}
\caption{System and stellar component parameters for J093010B}
\label{doubleparamtable}
\centering
\begin{tabular}{l | l l}
\hline\hline
 & Primary & Secondary \\
\hline
Orbital period (s) & \multicolumn{2}{|c}{$19674.47\pm0.03$} \\
Semi-major axis (R$_{\sun}$) & \multicolumn{2}{|c}{$1.665\pm0.012$} \\
Mass ratio & \multicolumn{2}{|c}{$0.397\pm0.006$} \\
System velocity (km~s\textsuperscript{-1}) & \multicolumn{2}{|c}{$-11.3\pm0.7$} \\
Angle of incl. (\degr) & \multicolumn{2}{|c}{$86\pm4$} \\
Kopal potential & \multicolumn{2}{|c}{$2.63\pm0.08$} \\
Filling factor & \multicolumn{2}{|c}{0.17} \\
Mass (M$_{\sun}$) & $0.86\pm0.02$ & $0.341\pm0.011$ \\
Radius (R$_{\sun}$) & $0.79\pm0.04$ & $0.52\pm0.05$ \\
Temperature (K) & \multicolumn{2}{|c}{$4700\pm50$} \\
Surface gravity & 4.58 & 4.53 \\
Bolometric luminosity & $6.20\pm0.05$ & $7.12\pm0.05$ \\
\hline
\end{tabular}
\end{table}

For both systems, distances were estimated by a similar method to that
outlined in \citeauthor{koo}: absolute bolometric magnitudes (outputs
from PHOEBE) were converted to absolute V magnitudes using bolometric
corrections as tabulated in \citet{flower}, and then combined using
the formula \citep{hilditch}
\begin{equation}M_{V,total}-M_{V,2}=-2.5\log_{10}(1+10^{-0.4(M_{V,1}-M_{V,2})}).\end{equation}
Apparent V magnitudes of the maximum brightness of the system were
determined from \citeauthor{koo}'s V light curves at their maxima for
J093010B; for J093010A the contribution of the third star had to be
removed first, using the component light ratios obtained during
analysis.  The formula \citep{hilditch}:
\begin{equation}V-A_V-M_{V,total}=5\log_{10}d-5\end{equation} was then
used to obtain distance estimates in parsecs (taking $A_V=0.03$).  For
J093010A, $d=78\pm3$~pc was found, and for J093010B, $d=73\pm4$~pc;
the ranges overlap substantially at 75--77~pc.  \citeauthor{koo} found
a consistent, if more uncertain value for J093010B of 77$\pm$9~pc, but
a shorter distance of 66$\pm$7~pc for J093010A, probably due in the
main to their handling of the third star's contribution to system
light.  The similarity of our distance estimates lends further support
to the reality of the association between the two binaries.

An approximate distance was also estimated for the fifth star: its
apparent V magnitude was found similarly to those for the binaries,
from its fractional light contribution to J093010A and from the V
light curves of \citeauthor{koo}.  Its absolute V magnitude was
assumed to be that of a typical mid-main-sequence star with solar
metallicity and $T=5000\pm100$~K (combining the two independent
effective temperature measurements obtained earlier), and was taken
from Dartmouth stellar evolution models
\citep{dotter}\footnote{http://stellar.dartmouth.edu/models/}.  The
formula above then gave $d=81^{+9}_{-8}$~pc for a 5~Gyr star, or
$d=83\pm9$~pc for 10~Gyr; both ranges are consistent with the
75--77~pc distance estimate found for the two binaries.  Although less
reliable than the binary distances, this finding adds some support to
the claim that the fifth star is physically associated with J093010.

\begin{figure}
\resizebox{\hsize}{!}{\includegraphics{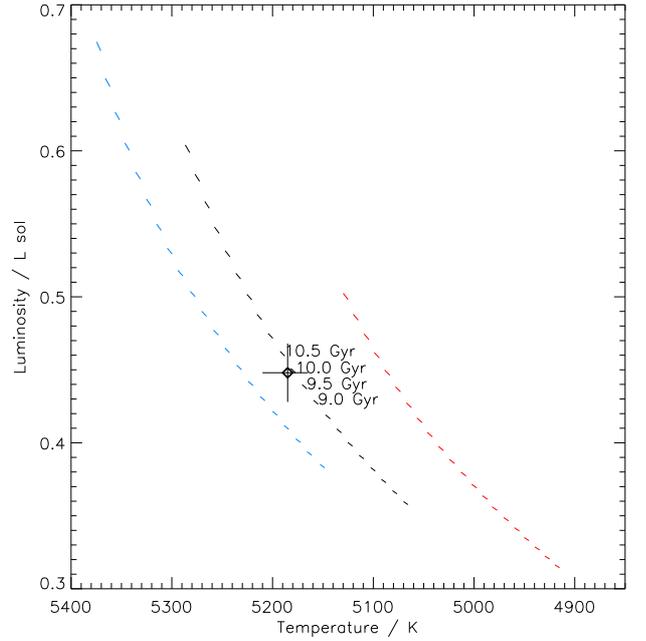}}
\caption{Temperature-luminosity plot for primary of J093010A binary,
  compared with Dartmouth stellar evolution model isochrones.  Each
  short curved line is a section of a model isochrone corresponding to
  masses between 0.829 and 0.845~M$_{\sun}$ i.e. the range of likely
  masses found from our modelling.  The sections in the middle (black)
  have solar metallicity; those on the right (red) have [Fe/H]=0.2
  and those on the left (blue) have [Fe/H]=$-$0.05.  Ages of four
  isochrones consistent with the primary's parameters are indicated.}
\label{templumprim}
\end{figure}

\begin{figure}
\resizebox{\hsize}{!}{\includegraphics{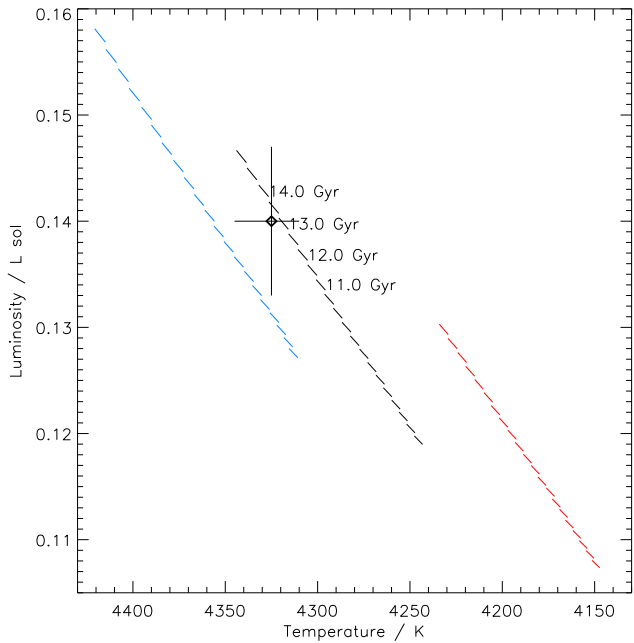}}
\caption{Temperature-luminosity plot for secondary of J093010A binary,
  compared with sections of model isochrones corresponding to
  masses between 0.667 and 0.681~M$_{\sun}$ (see caption to
  Fig.~\ref{templumprim} for further explanation).}
\label{templumsec}
\end{figure}

\begin{figure}
\resizebox{\hsize}{!}{\includegraphics{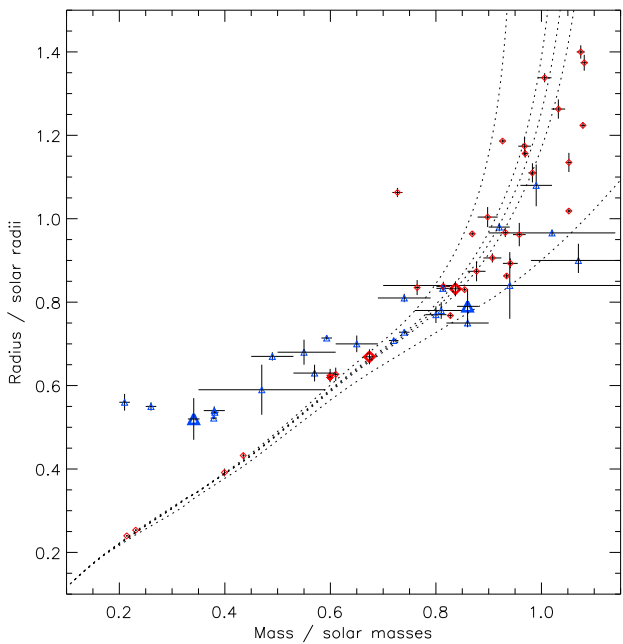}}
\caption{Mass-radius diagram for binary components of J093010,
  compared with Dartmouth model isochrones for solar metallicity
  (dotted lines, corresponding to 1, 8, 9, 10 and 14~Gyr from bottom
  to top), and with empirical masses and radii for collected detached
  (red diamonds) and contact (blue triangles) eclipsing binary
  components.  The binary components of J093010A are indicated by
  larger bold-face red diamonds, and those of J093010B by larger
  bold-face blue triangles.}
\label{massradplot}
\end{figure}

The parameters for J093010A are fully consistent with Dartmouth
models.  Plotting temperature against luminosity, each binary
component overlaps with the range of values expected for a star of its
mass, assuming solar metallicity (Figs.~\ref{templumprim} and
\ref{templumsec}).  A very tiny negative value of [Fe/H] is perhaps
implied for each, but nothing like the $-0.25$ found by
\citeauthor{koo}.  An optimal average age of $10.3\pm0.7$~Gyr is
indicated by this method.  Then, plotting masses against radii, both
components are consistent with solar metallicity isochrones between
around 8 and 10~Gyr (Fig.~\ref{massradplot}); this implies an optimal
average age of $9.2\pm1.0$~Gyr.

Fig.~\ref{massradplot} also shows that J093010A's binary components
behave very similarly to \citeauthor{torres}'s collection of low-mass
stars in detached eclipsing binaries with well-determined parameters
\citeyearpar{torres}.  The components of J093010B, however, do not sit
on a single isochrone, presumably because of their interaction in
contact: the primary has a smaller radius than its mass would imply,
assuming an age of 9 or 10 Gyr, while the secondary's radius is much
greater (and is inconsistent with any possible isochrone).  Comparison
with the parameters of short-period contact binary components
collected by \citet{stepien12}, supplemented by two determined in
\citet{lohr14}, suggests that this behaviour is entirely normal for
systems of this configuration: all the secondaries (those with masses
below 0.7~M$_{\sun}$) lie in this region of the mass-radius diagram,
with clearly inflated radii, while the primaries in most cases have
radii consistent with or even smaller than those expected from normal
stellar evolution models.

\section{Conclusion}

After our initial discovery of J093010 as a probable doubly eclipsing
quadruple system, and \citeauthor{koo}'s follow-up observations which
confirmed J093010A as a double-lined spectroscopic binary and revealed
the presence of a possible fifth component, we have carried
out further spectroscopic and photometric observations of the whole
system.  We have confirmed J093010B as a double-lined spectroscopic
binary also, and confirmed the association on the sky of the
fifth component within J093010A.  Radial velocities were measured for
the first time for J093010B, allowing a more reliable determination of
its mass ratio; the effective temperature of its primary, and of the
assumed tertiary in J093010A, were also estimated from their
disentangled individual spectra.  Reanalysed SuperWASP photometry
indicated the presence of a moving spot on one component of the binary
in J093010A, with an apparent speed relative to its host star of
approximately one rotation every two years.

After modelling both eclipsing systems we have obtained parameters for
them which agree with those of \citeauthor{koo} in some areas, but
disagree in several others, owing to different modelling approaches
and available data.  Specifically, we find a greater mass ratio for
J093010B on the basis of our new spectroscopic data, and different
radii for the binary components of J093010A as a result of our use of
the total-annular eclipses in its light curves, which also allowed
direct calculation of the third light contributed by the tertiary.
Both binaries probably possess at least one large spot, though further
observations would be required to constrain any spot properties
plausibly.

On the basis of our temperature estimates from the disentangled
component spectra, and from modelling, consistent distances for both
binaries of 73 and 78~pc are found.  The fifth non-eclipsing star
appears to have a temperature and spectrum similar to those of the
primaries in the two eclipsing binaries; a consistent distance is also
found for it if it is assumed to be a typical main-sequence star.  The
system velocities of both binaries, and the average radial velocity of
the fifth star, are all around $-11$ to
$-12$~km~s\textsuperscript{-1}.  These findings strongly support the
claim that both binaries, and very probably all five stars are
gravitationally bound in a single system.  The consistent angles of
inclination of the two binary systems also imply likely formation by
fragmentation from a single protostellar disk, without subsequent
significant disruption to their orbital plane.  The age of the system
is around 9--10~Gyr, and it has approximately solar metallicity.

This bright, close, highly unusual star system, containing a very
short-period contact eclipsing binary, and a triple system including a
low-mass detached eclipsing binary, would doubtless repay further
investigation.

\begin{acknowledgements}
  The WASP project is currently funded and operated by Warwick
  University and Keele University, and was originally set up by
  Queen's University Belfast, the Universities of Keele, St. Andrews
  and Leicester, the Open University, the Isaac Newton Group, the
  Instituto de Astrofisica de Canarias, the South African Astronomical
  Observatory and by STFC.  The William Herschel Telescope is operated
  on the island of La Palma by the Isaac Newton Group in the Spanish
  Observatorio del Roque de los Muchachos of the Instituto de
  Astrofisica de Canarias.  This work was supported by the Science and
  Technology Funding Council and the Open University, and accomplished
  with the help of the VO-KOREL web service, developed at the
  Astronomical Institute of the Academy of Sciences of the Czech
  Republic in the framework of the Czech Virtual Observatory (CZVO) by
  P.~Skoda and J.~Fuchs using the Fourier disentangling code KOREL
  devised by P.~Hadrava.
\end{acknowledgements}

\bibliographystyle{aa}
\bibliography{reflist}

\longtabL{1}{
\begin{landscape}
\begin{longtable}{l l l l l l l l l l l l}
\caption{Spectroscopic observations and heliocentric-corrected radial velocities for J093010.\label{J093010rvtable}}\\
\hline\hline
HJD & Exp. time /s & (A) Prim. RV /km~s\textsuperscript{-1} & $\delta$ Prim. RV & (A) Sec. RV & $\delta$ Sec. RV & (A) Ter. RV & $\delta$ Ter. RV & (B) Prim. RV & $\delta$ Prim. RV & (B) Sec. RV & $\delta$ Sec. RV \\ 
\hline
\endfirsthead
\caption[]{(continued)}\\
\hline\hline
HJD & Exp. & Prim. & $\delta$ Prim. & Sec. & $\delta$ Sec. & Ter. & $\delta$ Ter. & Prim. & $\delta$ Prim. & Sec. & $\delta$ Sec.\\ 
 & time & RV (A) & RV (A) & RV (A )& RV (A) & RV (A) & RV (A) & RV (B) & RV (B) & RV (B) & RV (B)\\
 & /s & /km~s\textsuperscript{-1} & /km~s\textsuperscript{-1} & /km~s\textsuperscript{-1} & /km~s\textsuperscript{-1} & /km~s\textsuperscript{-1} & /km~s\textsuperscript{-1} & /km~s\textsuperscript{-1} & /km~s\textsuperscript{-1} & /km~s\textsuperscript{-1} & /km~s\textsuperscript{-1}\\
\hline
\endhead
2456283.7679 & 5 & 87.5 & 1.0 & $-$132.3 & 1.2 & $-$11.8 & 0.5 & 28.9 & 1.5 & $-$116.4 & 1.5 \\
2456283.7704 & 30 & 90.1 & 0.7 & $-$132.5 & 1.1 & $-$11.1 & 0.6 & 40.4 & 1.5 & $-$138.1 & 1.5 \\
2456283.7710 & 30 & 88.6 & 1.6 & $-$133.0 & 1.3 & $-$11.1 & 0.3 & 39.2 & 1.5 & $-$143.4 & 1.5 \\
2456283.7716 & 30 & 89.7 & 0.6 & $-$130.9 & 0.5 & $-$11.3 & 0.9 & 44.5 & 1.5 & $-$141.7 & 1.5 \\
2456283.7749 & 180 & 88.3 & 0.5 & $-$133.3 & 0.9 & $-$11.4 & 0.6 & 51.5 & 1.5 & $-$162.8 & 1.5 \\
2456283.7773 & 180 & 88.7 & 0.8 & $-$133.2 & 0.3 & $-$11.0 & 0.3 & 52.6 & 1.5 & $-$179.0 & 1.5 \\
2456283.7796 & 180 & 88.4 & 0.4 & $-$133.3 & 0.3 & $-$10.98 & 0.25 & 55.9 & 1.5 & $-$180.8 & 1.5 \\
2456283.7853 & 360 & 89.1 & 0.4 & $-$133.0 & 0.8 & $-$11.2 & 0.8 & 69.2 & 1.5 & $-$203.4 & 1.5 \\
2456283.7897 & 180 & 89.0 & 0.5 & $-$133.4 & 1.2 & $-$11.5 & 0.4 & 71.1 & 1.5 & $-$219.4 & 1.5 \\
2456283.7921 & 180 & 88.7 & 0.3 & $-$133.0 & 0.9 & $-$10.58 & 0.14 & 74.1 & 1.5 & $-$237.1 & 1.5 \\
2456283.7979 & 360 & 86.8 & 1.0 & $-$134.2 & 2.0 & $-$13.0 & 1.6 & & & & \\
2456292.7661 & 60 & 58.01 & 0.18 & $-$102.3 & 1.5 & $-$12.35 & 0.18 & $-$64.6 & 1.5 & 124.7 & 1.5 \\
2456292.7671 & 60 & 60.3 & 0.7 & $-$101 & 3 & $-$12.3 & 0.6 & $-$63.3 & 1.5 & 125.4 & 1.5 \\
2456292.7680 & 60 & 59.0 & 0.6 & $-$99.6 & 1.3 & $-$11.8 & 0.5 & $-$66.7 & 1.5 & 132.3 & 1.5 \\
2456292.7710 & 10 & 58.8 & 1.6 & $-$98 & 5 & $-$13.3 & 1.1 & $-$79.4 & 1.5 & 147.3 & 1.5 \\
2456292.7743 & 10 & 55.9 & 1.3 & $-$102.5 & 2.2 & $-$13.1 & 0.3 & $-$84.1 & 1.5 & 163.5 & 1.5 \\
2456292.7758 & 60 & 59.4 & 0.8 & $-$106.4 & 0.8 & $-$13.39 & 0.19 & $-$88.6 & 1.5 & 170.3 & 1.5 \\
2456292.7767 & 60 & 59.7 & 0.8 & $-$106.1 & 2.5 & $-$13.1 & 0.5 & $-$88.7 & 1.5 & 173.2 & 1.5 \\
2456292.7777 & 60 & 59.33 & 0.12 & $-$107.54 & 0.18 & $-$14.3 & 0.7 & $-$93.9 & 1.5 & 176.0 & 1.5 \\
2456292.7822 & 360 & 61.2 & 0.5 & $-$106.6 & 0.4 & $-$13.2 & 0.4 & $-$100.4 & 1.5 & 197.7 & 1.5 \\
2456292.7866 & 360 & 62.5 & 0.6 & $-$107.6 & 0.5 & $-$13.2 & 0.5 & $-$107.0 & 1.5 & 213.8 & 1.5 \\
2456292.7911 & 360 & 63.8 & 0.4 & $-$110.5 & 0.7 & $-$13.16 & 0.18 & $-$109.8 & 1.5 & 229.5 & 1.5 \\
2456292.7948 & 60 & 67.2 & 1.5 & $-$113.0 & 0.6 & $-$16.8 & 1.8 & $-$118.7 & 1.5 & 237.3 & 1.5 \\
2456292.7957 & 60 & 68.8 & 2.3 & $-$110.2 & 0.6 & $-$10.9 & 0.7 & $-$116.4 & 1.5 & 239.8 & 1.5 \\
2456292.7967 & 60 & 69.3 & 0.8 & $-$112.5 & 0.9 & $-$12.8 & 0.9 & $-$114.4 & 1.5 & 246.6 & 1.5 \\
2456292.7992 & 300 & 66.4 & 0.7 & $-$112.4 & 0.8 & $-$13.6 & 0.4 & $-$116.0 & 1.5 & 247.1 & 1.5 \\
2456292.8029 & 300 & 67.7 & 0.4 & $-$112.2 & 0.9 & $-$12.83 & 0.13 & $-$119.0 & 1.5 & 247.5 & 1.5 \\
2456292.8066 & 300 & 68.49 & 0.06 & $-$114.9 & 0.9 & $-$13.1 & 0.3 & $-$117.5 & 1.5 & 247.6 & 1.5 \\
2456294.7665 & 60 & $-$91.7 & 0.8 & 96.0 & 1.7 & $-$9.05 & 0.13 & 67.1 & 1.5 & $-$209.4 & 1.5 \\
2456294.7674 & 60 & $-$91.4 & 1.4 & 96.3 & 2.2 & $-$9.1 & 0.3 & 66.6 & 1.5 & $-$196.4 & 1.5 \\
2456294.7684 & 60 & $-$93.6 & 0.6 & 94.4 & 1.1 & $-$9.6 & 0.5 & 64.0 & 1.5 & $-$197.0 & 1.5 \\
2456294.7707 & 120 & $-$94.4 & 0.6 & 94.9 & 0.9 & $-$10.25 & 0.10 & 67.8 & 1.5 & $-$178.5 & 1.5 \\
2456294.7724 & 120 & $-$94.4 & 0.7 & 95.9 & 0.3 & $-$9.67 & 0.15 & 58.6 & 1.5 & $-$179.1 & 1.5 \\
2456294.7740 & 120 & $-$95.4 & 0.4 & 95.7 & 0.4 & $-$10.40 & 0.06 & 56.1 & 1.5 & $-$171.7 & 1.5 \\
2456294.7783 & 110 & $-$95.59 & 0.14 & 97.0 & 1.2 & $-$10.07 & 0.21 & 45.0 & 1.5 & $-$143.4 & 1.5 \\
2456294.7799 & 110 & $-$97.0 & 0.4 & 96.4 & 1.1 & $-$10.8 & 0.5 & 37.4 & 1.5 & $-$143.4 & 1.5 \\
2456294.7814 & 110 & $-$96.6 & 0.5 & 96.6 & 1.2 & $-$10.22 & 0.20 & 37.4 & 1.5 & $-$126.2 & 1.5 \\
2456294.7829 & 110 & $-$97.8 & 0.3 & 96.7 & 0.3 & $-$10.69 & 0.09 & 32.7 & 1.5 & $-$128.3 & 1.5 \\
2456294.7845 & 110 & $-$97.2 & 0.4 & 98.1 & 0.6 & $-$10.2 & 0.3 & 31.0 & 1.5 & $-$108.3 & 1.5 \\
2456294.7860 & 110 & $-$98.2 & 0.5 & 96.9 & 1.5 & $-$10.85 & 0.08 & 19.9 & 1.5 & $-$95.7 & 1.5 \\
2456294.7885 & 110 & $-$99.0 & 0.4 & 96.5 & 0.7 & $-$11.03 & 0.12 & 13.6 & 1.5 & $-$76.9 & 1.5 \\
2456294.7901 & 110 & $-$99.2 & 1.0 & 97.3 & 1.0 & $-$11.02 & 0.24 & 13.5 & 1.5 & $-$73.1 & 1.5 \\
2456294.7916 & 110 & $-$98.8 & 0.4 & 97.8 & 1.0 & $-$10.53 & 0.21 & 7.9 & 1.5 & $-$56.5 & 1.5 \\
2456294.7931 & 110 & $-$100.02 & 0.16 & 96.9 & 0.4 & $-$11.25 & 0.24 & $-$1.4 & 1.5 & $-$41.4 & 1.5 \\
2456294.7947 & 110 & $-$99.4 & 0.4 & 98.3 & 0.8 & $-$10.61 & 0.22 & $-$2.3 & 1.5 & $-$35.0 & 1.5 \\
2456294.7962 & 110 & $-$100.3 & 0.4 & 97.7 & 1.4 & $-$11.39 & 0.20 & $-$5.2 & 1.5 & $-$29.5 & 1.5 \\
2456294.7986 & 60 & $-$99.7 & 0.6 & 100.5 & 1.3 & $-$10.5 & 0.6 & $-$16.1 & 1.5 & $-$3.7 & 1.5 \\
2456294.7996 & 60 & $-$100.5 & 1.1 & 98.7 & 1.6 & $-$10.8 & 0.3 & $-$10.2 & 1.5 & $-$20.6 & 1.5 \\
2456294.8005 & 60 & $-$102.0 & 0.7 & 99.6 & 0.5 & $-$11.9 & 0.3 & $-$17.3 & 1.5 & $-$5.8 & 1.5 \\
\hline
\end{longtable}
\end{landscape}
}

\end{document}